# Contribution to the study of sub-bandgap photon absorption in quantum dot InAs/AlGaAs intermediate band solar cells

Juan Villa, Iñigo Ramiro, José María Ripalda, Ignacio Tobías,

Pablo García-Linares, Elisa Antolín and Antonio Martí

*Abstract*— Intermediate band solar cells (IBSCs) pursue the increase in efficiency by absorbing below-bandgap energy photons while preserving the output voltage. Experimental IBSCs based on quantum dots have already demonstrated that both below-bandgap photon absorption and the output voltage preservation, are possible. However, the experimental work has also revealed that the below-bandgap absorption of light is weak and insufficient to boost the efficiency of the solar cells. The objective of this work is to contribute to the study of this absorption by manufacturing and characterizing a quantum dot intermediate band solar cell with a single quantum dot layer with and without light trapping elements. Using one-dimensional substrate texturing, our results show a three-fold increase in the absorption of below bandgap energy photons in the lowest energy region of the spectrum, a region not previously explored using this approach. Furthermore, we also measure, at 9K, a distinguished split of quasi-Fermi levels between the conduction and intermediate bands, which is a necessary condition to preserve the output voltage of the cell.

## I. INTRODUCTION

The intermediate band solar cell (IBSC) concept pursues the increase in the efficiency of solar cells thanks to the absorption of below bandgap energy photons by means of an "intermediate" electronic band properly located in-between the conduction and valence bands (CB and VB) [1]. As detailed in Fig. 1, the existence of this intermediate band (IB) splits the total bandgap of the semiconductor $E_G$ into two smaller gaps, $E_H$ and $E_L$. In addition to the absorption of below bandgap energy photons, represented by arrows (1) and (2) in the figure, the output voltage of the cell has to be preserved, meaning that it should not be limited by any of the smaller gaps, $E_H$ nor $E_L$, but only by the total bandgap $E_G$. This is ultimately possible thanks to the fact that the electron concentration in each of the bands is described by its own quasi-Fermi level: $\mu_C$ for electrons in the CB, $\mu_I$ for electrons in the IB, and $\mu_V$ for electrons in the VB. In this configuration, the output voltage of the cell, $V$, is related to the CB and VB quasi-Fermi levels by:
$$eV = \mu_C - \mu_V \qquad (1)$$
$e$ being the charge of the electron in absolute value.

Several strategies have been proposed to implement this concept and the interested reader can consult publications such as [2][3]. In this respect, the framework in which our work fits is that in which the IBSC is developed by implementing InAs/(Al)GaAs quantum dots (QDs). Using this approach, the IB arises from the energy states associated to the electrons confined in the CB of the QDs as symbolically illustrated in Fig. 1. The implementation of the IBSC by means of QDs has allowed several of the principles of operation of the IBSC to be demonstrated such as the absorption of two below-bandgap energy photons to produce one electron hole pair [4] and the preservation of the output voltage of the cell [5]. Blokhin et al. [6] have reported efficiencies above 18 % for this system. Sablon et al. [7] have also reported efficiencies in this range and above 21 % under concentrated light (40-90 suns). Bailey et al. [8] have reported the highest open-circuit voltage (near 1V) and efficiencies above 13 %.

Despite this, the efficiency of the device remains low to a large extent, in our opinion, because the absorption of light by the QDs remains low [9]. In this respect, Pusch et al. [10] have discussed the impact of low sub-gap absorptivity in the efficiency of an IBSC. Either way, the problems associated to low photon absorptivity precedes other problems, such as the ones derived from poor material quality and carrier lifetime, since, even if good lifetimes are achieved, it is impossible to

Manuscript received July 18th 2020. This work was supported in part by the Project 2GAPS (TEC2017-92301-EXP) and MULTIPLIER (RTI2018-096937-B-C21) funded by the Spanish Ministerio de Ciencia, Innovación y Universidades, the Project MADRID-PV2-CM (P2018/EMT-4308) funded by the Comunidad de Madrid supported with FEDER funds. E. A. is funded by a Ramón y Cajal Fellowship from the Spanish Ministry of Science (RYC-2015-18539). are preferred in the author field, but are not required. Put a space between authors' initials.

J.V Author was at the Instituto de Energía Solar of the Universidad Politécnica de Madrid during this work. Now is at the Instituto Astrofísico de Canarias.

I. R., I.T, P.G.L, E.A and A. M Authors are with Instituto de Energía Solar of the Universidad Politécnica de Madrid.

J.M.R. Author is with Instituto de Micro y Nanotecnología of the Centro Nacional de Microelectrónica-CSIC.

I.R is the corresponding author (i.ramiro@ies.upm.es).

make a good solar cell without good absorption.

Some attempts have already been made to increase absorption in the QDs by incorporating light-management structures to optoelectronic devices (solar cells and infrared photodetectors) involving QDs. In the case of infrared photodetectors, for example, Chang et al. [11] exploited surface plasmon enhanced resonance to show an increase in the QD absorption by a factor 11, which, in the notation used in Fig. 1 would correspond to transition "2". The device used in these experiments contained 30 layers of QDs. However, the metallic two-dimensional hole array (2DHA) introduced on the front surface of the device to achieve the plasmonic resonance, also introduced a 85 % shadowing factor which, although promising, would still make the use of the structure unpractical for photovoltaic applications. Furthermore, plasmonic resonance has a strongly monochromatic effect, which could make it difficult to exploit in solar cells where enhancement in photon absorption in a broader spectrum range is sought. Nevertheless, Liu et al. [12] optimized the 2DHA for photovoltaic applications and found a 6 X photocurrent enhancement, but with a different impact on the performance of the devices depending on the number of QD layers incorporated. To avoid front shadowing, Feng Lu et al. [13] incorporated plasmon resonant structures consisting of Ag nanoparticles at the back of a quantum dot solar cell incorporating 10 QD layers and obtained an increase of around 5 % in the photocurrent. In this case, the resonance acted on the transition labeled "2" in Fig. 1.

In the context of far-field optics, simulations by Aho et. al. [14] have predicted a 12-fold increase in comparison with an structure without a reflector if a pyramidal grating were used as a back reflector (this would correspond to approximately a 6-fold factor in comparison with a structure with a flat back reflector). This improvement could increase to a 27-fold factor ($\sim 13-14$ -fold in comparison with a structure with a back reflector) if a grating is used both at the front and the back of the cell [15]. This improvement would correspond to photon wavelengths corresponding to transitions labelled "1" in Fig. 1. Previously, Elsehraway et al. [16] had anticipated a 4-fold improvement at least (in comparison with a structure without a reflector).

In the framework of geometrical optics, Smith et al. [17] compared solar cells incorporating 10 layers of QDs, manufactured using the epitaxial lift-off (ELO) technique, with a flat and textured back reflector, and obtained an increase of 30 % in below bandgap photocurrent in the textured sample with respect to the flat reflector one. The improvement was attributed to an increased generation rate related to transition "1" in Fig. 1. More recently Shoji et al. [18], using solar cells involving 50 layers of QDs and a chemical texturing of the back of the cell found a 2.4 X increase in photocurrent when compared with an untextured reference. The improvement was also attributed to an increased carrier generation associated to transition "1".

This work attempts to contribute further to the study of the enhancement of photon absorption that can be expected in quantum-dot, intermediate-band solar cells (QD-IBSCs) using conventional light trapping techniques. Following [18], we will use also a semi-insulating GaAs substrate to avoid parasitic absorption. However, we will use a single QD layer in our devices, so that the results can be uniquely associated to the structural properties of this single layer, and a one-dimension textured pattern at the back of the cell, instead of a random texture, to facilitate the comparison between theory and experimental results. To ensure the reproducibility of the back reflectivity of our cells, the back of all samples is left to the air. As a result, we shall see that our approach provides insight into the photon absorption related to transition "2", a region of the spectrum not previously explored using this technique, as well as experimental evidence of the quasi-Fermi level split also illustrated in Fig. 1.

Our paper is structured as follows: Section II describes the solar cell architecture that we use, which facilitates the study of light trapping in the infrared, the region of the spectrum where the QDs absorb below bandgap light. We also illustrate the difficulty in studying the absorption of below-bandgap energy photons by means of standard solar cell current-voltage characteristics (section III) which can lead to misleading results, and the quantum efficiency of the cells. Finally, in section IV we describe how electroluminescence can be used to obtain evidence of the split of quasi-Fermi levels between the conduction and the intermediate bands, $\mu_C$ and $\mu_I$.

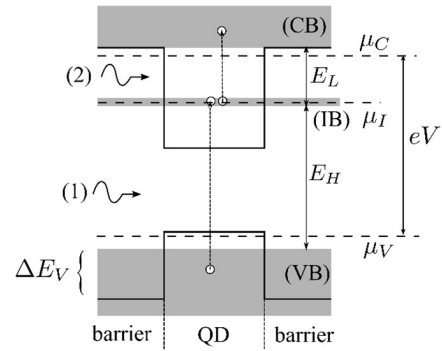

Fig. 1. Simplified energy band diagram of QD-based IBSC showing the optical transitions and energy gaps of interest. Photon (1) is shown promoting a transition from the valence band (VB) to the intermediate band (IB) and photon (2) is shown promoting a transition from the IB to the conduction band (CB). $E_L$ indicates the gap between the IB and the CB and $E_H$ indicates the gap between the VB and the IB. $\Delta E_V$ indicates the valence band offset (the energy difference between the valence band of the barrier material and the first confined states in the valence band of the QD); $\mu_C$, $\mu_I$ and $\mu_V$ are the electrochemical potentials (or quasi-Fermi levels) for electrons in the CB, IB and VB. $V$ is the output voltage of the cell.

II. DEVICE STRUCTURES

Fig. 2 shows a simplified version of the solar cell structure used in this work. Only the main semiconductor layers relevant to the operation of the solar cell have been represented. A detailed description of the solar cells structure, containing all the semiconductor layers (contact, damping layers, etc…) can be found in Annex I. We will now describe the main features of this structure.

First, the solar cell is grown on a semi-insulating (SI) GaAs substrate. The reason is that we want below bandgap energy photons to reach the back surface of the cell by minimizing parasitic free carrier absorption [19], which increases when the substrate is doped. Therefore, in the context of the light trapping analysis, "below bandgap energy photons" will mean photons with energy lower than the GaAs gap ($\approx 1.42$ eV at room temperature) [20] since these are the only photons capable of reaching the back of the cell. This energy is sufficient to pump (Fig. 1) electrons from the VB to the IB (since, as we shall see, $E_H \approx 1.15$ eV) as well as from the IB to the CB (since, as we shall find out, $E_L \approx 0.40$ eV).

Second, we access the back contact of the cell laterally through the $n^+$-buffer layer. The motivation for this approach is that we want to have the rear of the substrate free from any metallization so that we can keep this surface polished, leading to the sample we shall call a POL cell (Fig. 2a) or we can texture it, leading to the sample we shall call a TEX cell (Fig. 2b). In our cell, texturing takes place according to a one-dimensional (1D) U-like pattern (Fig. 2d) of around 30 µm in periodicity and 4.5 µm in depth. These grooves at the back are created with a linear pattern mask, using standard photolithography with a positive photoresist and controlled HF:HNO$_3$:acetic acid wet chemical etching. For our studies, we have used one POL and one TEX samples (see ANNEX I).

In both samples, POL and TEX, the back of the cell is in contact with air thanks to the solar cell assembly described in Annex I (Fig. 6c). Having air at the back avoids uncertainty related to the reflection of light at this surface since in both structures we have a semiconductor-air interface. However, in the POL sample, light coming from the sun escapes through the back of the cell and the bandgap energy photons below have an optical path, $l$, $l$ being the effective thickness of the QD layer. On the other hand, in the TEX sample, light is scattered thanks to the texturing, and the effective optical path is increased. For an ideal Lambertian scatterer, the optical path of below bandgap energy photons would increase [21] to $4n^2l \approx 50l$, $n$ being $\approx 3.6$, the refraction index of the semiconductor. Our texturing is, however, one-dimensional and therefore not ideal, and shorter in optical length. As expected, its upper limit enhancement is equal to $\pi n l \approx 11.3l$ [22].

Third, our cell consists of one QD layer only. This layer is sandwiched between a 600 nm thick undoped Al$_{0.35}$Ga$_{0.65}$As layer. We use one single QD layer because we want our results to be attributed exclusively to the properties of this one layer and therefore avoiding a complex analysis resulting from, for example, an inhomogeneous electron population or carrier collection efficiency between the different layers of QDs. QDs have been grown in the Stranski-Krastanov growth mode [23] by molecular beam epitaxy (MBE), which creates a thin In(Ga,Al)As quantum well layer underneath the QDs known as a wetting layer (WL). Expected QD density per unit of area is $3 \times 10^{10}$ cm$^{-2}$. Fig. 2c shows a HRTEM picture of the QDs grown, revealing they have a pyramidal shape of around 5 nm in height and a base of around 20 nm. During growth, QDs have been Si doped nominally to $1.5 \times 10^{10}$ cm$^{-2}$ in order to partially populate the confined electron states (the IB). This is necessary since, ideally, this band should have both empty and occupied states to receive electrons from the VB and to supply them to the CB [24]. However, this partial filling is difficult to control and our experimental results, described later, should be able to tell us whether this goal has been achieved or not.

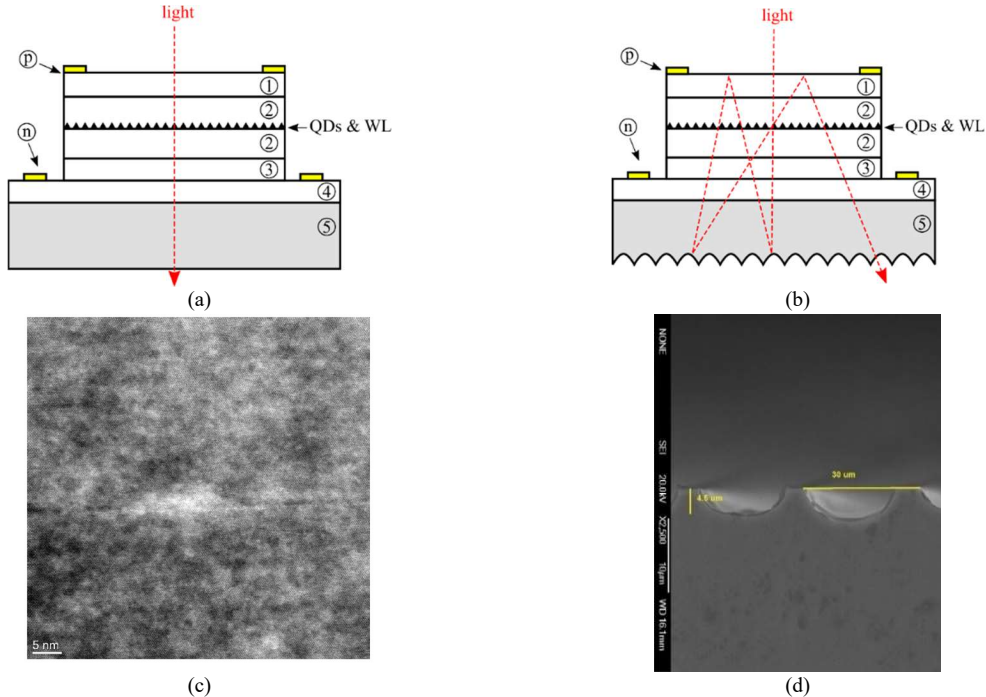

Fig. 2. QD-based IBSC structures used in this work: (a) with the back polished; (b) with the back textured following a U pattern. Fresnel losses for the light are not represented. (c) TEM micrograph of an InAs QD in Al$_{0.35}$GaAs$_{0.65}$ barrier; (d) SEM image of the resulting one dimensional U pattern at the back of the cell. In both figures; 1: p-AlGaAs; 2:undoped AlGaAs; 3: n-AlGaAs, 4: n+ GaAs buffer; 5: SI-GaAs substrate. p and n contacts are also indicated. QDs & WL stands for Quantum Dots and Wetting Layer. More details of the semiconductor structure can be found in Annex I.

### III. CURRENT-VOLTAGE AND QUANTUM EFFICIENCY CHARACTERISTICS

The current-voltage characteristics of the POL and the TEX solar cells under AM1.5 Global-1000 Wm$^{-2}$ illumination [25] are detailed in Fig. 3a. The inset show the dark current–voltage characteristics. These characteristics reveal an increased short-circuit current of the POL cell when compared to the TEX cell which could lead us to the conclusion that below bandgap light trapping is better achieved in the POL cell than in the TEX cell. However, this conclusion will prove to be wrong after the analysis of the quantum efficiency of the cells provided in Fig. 3b. In this plot we observe that there is no significant photo-response for photons with energy lower than the GaAs gap (1.42 eV ), the only photons that according to our discussion in the section before can reach the back surface of the cell. Therefore, we conclude that the higher short-circuit current of the POL sample with respect to the TEX sample comes from the accidental better QE response of the POL cell to photons with energy between the GaAs gap and Al$_{0.35}$Ga$_{0.65}$As gap (1.90 eV [26]) together with a better response to ultraviolet energy photons (energy higher than 2.5 eV).

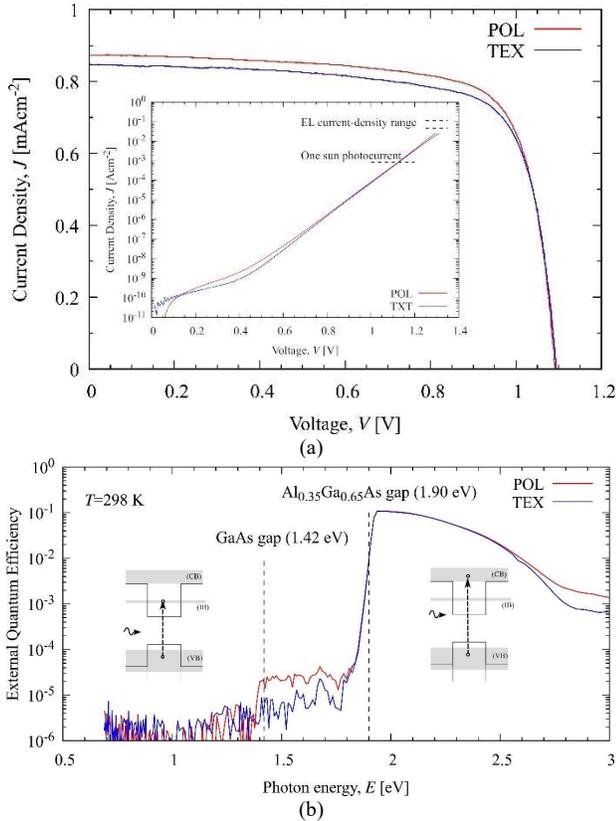

Fig. 3. (a) Current-voltage characteristic of the POL and TEX solar cells at 298 K and under AM1.5G illumination. The inset shows the dark current-voltage characteristics indicating for reference some of the current levels used in the experiments; (b) External quantum efficiency.

The photo-response of the cell to photons with energy higher than the GaAs gap stems from the presence of GaAs in several of the layers of the cell structure: SI-substrate, $n^+$ buffer layer, as well as a top $p^+$ contact layer (illustrated in Annex I, Fig. 6a). The possible contribution of the substrate to the photocurrent of an IBSC was discussed in [27].

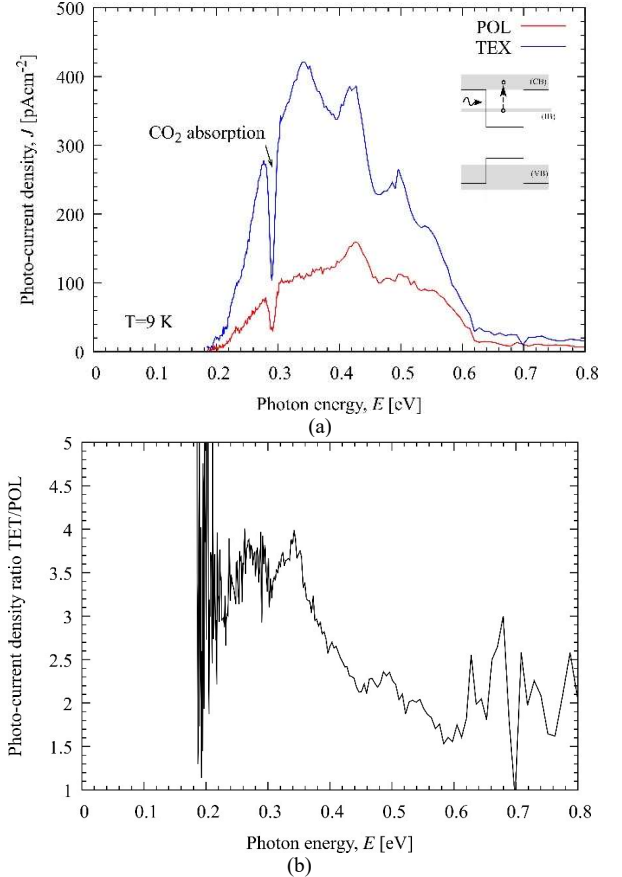

Fig. 4. (a) Spectral photocurrent for the POL and TEX samples (for reference, we estimate an excitation power density of 0.3 mWcm$^{-2}$ @ 0.4 eV and for an excitation bandwidth of 75 nm.); (b) Ratio between the spectral photocurrents of the POL and TEX samples.

As we have just discussed, the QE in Fig. 3b does not show any photo-response to photons with energy lower than the GaAs gap (1.42 eV). Although, in an ideal IBSC, two below bandgap photons are needed to promote an electron from the VB to the CB and contribute to the photo-current, it is common though that one photon alone (with energy higher than $E_H$) gives rise to a photo-response at room temperature. The reason is that, after being optically excited from the VB to the IB, electrons can escape from the IB to the CB via thermal activation [28]. Therefore, the absence of a photo-response from the IB associated to the QD layer is unexpected. Our interpretation of this result is that the intentional doping of the QDs in these samples has completely filled the IB with electrons and, therefore, electron transitions from the VB to the IB are not possible because there are no empty states in the IB to receive electrons. However, if this is true, transitions from the IB to the CB due to photon absorption should be possible and, in fact, they are, as demonstrated by the spectral photo-current measurements shown in Fig. 4a. These measurements are taken at a low temperature (9 K) because thermal noise masks the signal at room temperature. Notice that results in Fig. 4a are not quantum efficiency measurements. Nevertheless, measuring the spectral photocurrent for both the POL and TEX samples under the same experimental conditions is sufficient to determine the increase in the relative increase in the optical path for below the GaAs bandgap energy photons. This is done in Fig. 4b, which reveals an increase in the optical length in the TEX sample of around 3.5 times that of the POL sample for photons with energy of around 0.25-0.35 eV. This result is in agreement with ray-tracing simulations where we obtain an equivalent optical length of 3.4 (Annex III). It is important to emphasize that, as long as we are within the framework of geometric optics – which given the dimensions of our geometries and photon energies involved, is a valid approximation (see Annex III) – this number is independent of the photon wavelength.

## IV. ELECTROLUMINESCENCE

The measurement of the light emitted by the cells when biased with forward current (electroluminescence) can provide insight into the split of quasi-Fermi levels between the conduction band ($\mu_C$) and the intermediate band ($\mu_I$) [29]. This approach could complement other contactless approaches developed based on photoluminescence [30]. In this respect, Fig. 5 plots the electroluminescence of the POL and TEX cells at 9 K, as measured from the front of the cells, and for several current biasing conditions. Several peaks appear and are identified as follows:

- Peak A ($E_H \approx 1.15$eV): We interpret that it corresponds to transition from the lowest energy state in the IB to the valence band and reveals the value of the gap $E_H$ (Fig. 1).
- Peak B ($E_S \approx 1.47$eV): We think it is plausible that this peak corresponds to the electroluminescence from the GaAs. At 9 K, SI- GaAs has a bandgap of 1.52 eV [26]. However, our structure (Fig. 6a) contains two highly doped GaAs layers (the $n^+$-buffer layer and the $p^+$-top contact layer). The lower value of the energy of this emission peak when compared to the gap of SI-GaAs could be due to bandgap narrowing effects resulting from the high doping of these layers [31][32].
- Peak D ($E_G \approx 2.00$ eV) matches the value of $Al_{0.35}Ga_{0.65}As$ gap [26]. Together with the previous value, $E_H \approx 1.15$eV, this leads us to conclude that, for zero valence band offset in the QD ($\Delta E_V = 0$), $E_L \approx 0.85$eV. The peak of around 0.40 eV in Fig. 4a would suggest instead that $E_L \approx 0.40$eV and a valence band offset of $\Delta E_V \approx 0.45$eV (see Fig. 1). The value of 0.4 eV must be understood as representative of some middle position of energy states at the IB whose bandwidth could extend $\pm 200$ meV around this position.
- Peak C ($E_W \approx 1.63$eV) has a lower value than $Al_{0.35}Ga_{0.65}As$ gap and we attribute it either to the wetting layer or to the existence of a higher energy bound state in the QD. Either way, the origin of this peak will be irrelevant in the discussion that follows.

We shall now pay our attention to the relative intensity of peaks A and D, located at the extreme of the electroluminescence energy bandwidth response. Assuming a simplified model in which the quasi-Fermi levels are flat, the intensity of the electroluminescence peak associated to peaks A and D, which we will call $P_{POL}(E_H)$ and $P_{POL}(E_G)$ respectively, for the POL sample, are given approximately by:

$$P_{POL}(E_G) \approx D(E_G)a_{POL}(E_G)K_{POL}\exp\left[\frac{(\mu_C - \mu_V)_{POL}}{kT}\right] \quad (2)$$

$$P_{POL}(E_H) \approx D(E_H)a_{POL}(E_H)K_{POL}\exp\left[\frac{(\mu_I - \mu_V)_{POL}}{kT}\right] \quad (3)$$

where: $\mu_C$, $\mu_V$, $\mu_I$ are the electro-chemical potentials of the electrons in the conduction, valence and intermediate band respectively; $k$ is the Boltzmann constant, $T$ is the absolute temperature of the cell; $D(E)$ is the detectivity of the detector, that depends on the photon energy $E$ being detected and not on the type of sample being tested; $K_{POL}$ includes all constants that do not depend on the photon energy associated to the POL sample (shadowing of the metal grid, area, misalignments of the optical system…) and $a_{POL}(E)$ is the absorptivity of sample POL to photons of energy $E$ when incident from the front. We recall that the appearance of the absorptivity $a_{POL}(E)$ in these equations, as well as the exponential terms involving the quasi-Fermi level split, stems from the application of detailed balance theory, proposed in [33] and revised by some of us in [34][35]. In this respect, we recall that the absorption coefficient considered by the theory, and related to this absorptivity, must involve only the corresponding band to band optical transitions and not free-carrier absorption.

We can formulate similar equations for the TEX sample:

$$P_{TEX}(E_G) \approx D(E_G)a_{TEX}(E_G)K_{TEX}\exp\left[\frac{(\mu_C - \mu_V)_{TEX}}{kT}\right] \quad (4)$$

$$P_{TEX}(E_H) \approx D(E_H)a_{TEX}(E_H)K_{TEX}\exp\left[\frac{(\mu_I - \mu_V)_{TEX}}{kT}\right] \quad (5)$$

where, now, $K_{TEX}$ involves all constants that do not depend on the photon energy associated to the TEX sample and $a_{TEX}(E)$ is the absorptivity of the TEX sample to photons of energy $E$ when incident from the front.

From the data in Fig. 5 we can calculate the value of the ratio $R$ defined as:

$$R = \frac{P_{TEX}(E_H)P_{POL}(E_G)}{P_{TEX}(E_G)P_{POL}(E_H)} \quad (6)$$

Using the results in Eqs. (2) to (5) we also find that:

$$R = \frac{a_{TEX}(E_H)a_{POL}(E_G)\exp\left[\frac{(\mu_I - \mu_C)_{TEX}}{kT}\right]}{a_{POL}(E_H)a_{TEX}(E_G)\exp\left[\frac{(\mu_I - \mu_C)_{POL}}{kT}\right]}$$
$$= \frac{a_{TEX}(E_H)\exp\left[\frac{(\mu_C - \mu_I)_{POL}}{kT}\right]}{a_{POL}(E_H)\exp\left[\frac{(\mu_C - \mu_I)_{TEX}}{kT}\right]} \quad (7)$$

where we have used that for photons with energy equal to $E_G$, since they do not reach the back of the cell, $a_{POL}(E_G) = a_{TEX}(E_G)$.

To illustrate the implications of this result, let us assume that $(\mu_C - \mu_I)_{POL} = (\mu_C - \mu_I)_{TEX}$. Note that this assumption includes the case $(\mu_C - \mu_I)_{POL} = (\mu_C - \mu_I)_{TEX} = 0$. Then,

$$R = \frac{a_{TEX}(E_H)\exp\left[\frac{(\mu_I - \mu_C)_{TEX}}{kT}\right]}{a_{POL}(E_H)\exp\left[\frac{(\mu_I - \mu_C)_{POL}}{kT}\right]} = \frac{a_{TEX}(E_H)}{a_{POL}(E_H)} \quad (8)$$

The measured values of $R$ are shown in Table I and reveal that, assuming $(\mu_C - \mu_I)_{POL} = (\mu_C - \mu_I)_{TEX}$ would lead us to absorptive ratios of above 20 between the TEX and POL samples for below bandgap energy photons, significantly higher than the values measured using direct spectral photocurrent measurements (Fig. 4), in agreement with our modelling (Annex III), and even well beyond the theoretical limit (11.3) anticipated in Section II. Our conclusion, in this respect, is that we must have $(\mu_C - \mu_I)_{POL} \neq (\mu_C - \mu_I)_{TEX}$ and, in particular $(\mu_C - \mu_I)_{POL} \neq 0$ or $(\mu_C - \mu_I)_{TEX} \neq 0$ which is also one of the working hypotheses of the IBSC theory [1]. Also, since $R > 1$, we find that the split in cuasi-Fermi levels in the POL sample is larger than in the TEX sample. We think this is accidental and perhaps attributed to the more complex processing of the TEX sample that might cause an extra component of non-radiative recombination that could also support the poorer quantum efficiency response in the short-wavelength region of the TEX sample we showed in Fig. 3,b.

TABLE I. ABSORPTIVITY RATIO BETWEEN THE TEX AND POL CELLS FOR BELOW BANDGAP ENERGY PHOTONS ASSUMING $(\mu_C - \mu_I)_{POL} = (\mu_C - \mu_I)_{TEX}$.

| Current bias | $\dfrac{a_{TEX}(E_H)}{a_{POL}(E_H)}$ |
|---|---|
| 45.2 mAcm$^{-2}$ | 20.4 |
| 67.8 mAcm$^{-2}$ | 24.9 |
| 113.1 mAcm$^{-2}$ | 50.1 |

## V. CONCLUSIONS

We have manufactured a QD-IBSC structure that allows its back to be processed in order to implement light trapping architectures such as texturing in order to study their impact. This structure is grown on an SI substrate in order to minimize free-carrier absorption and allow low energy photons to reach this rear side. The price to pay is that the electric current has to be extracted laterally. We have applied the structure to the experimental study of the light confinement provided by one-dimensional texturing finding an increase in the optical path of around 3.5 times, for photon energies in the 0.20-0.35 eV range (lowest gap of the IBSC), with respect to a sample with its rear side polished. We have also shown that, because below bandgap photon absorption with one QD layer produces negligible effects in the short-circuit current of the solar cells, this impact is more conveniently analyzed through quantum efficiency measurements. Finally, we have illustrated that electroluminescence measurements allows the energy levels relevant to the solar cell performance to be identified and how they could also provide evidence of the existence of quasi-Fermi level split between the IB and the CB.

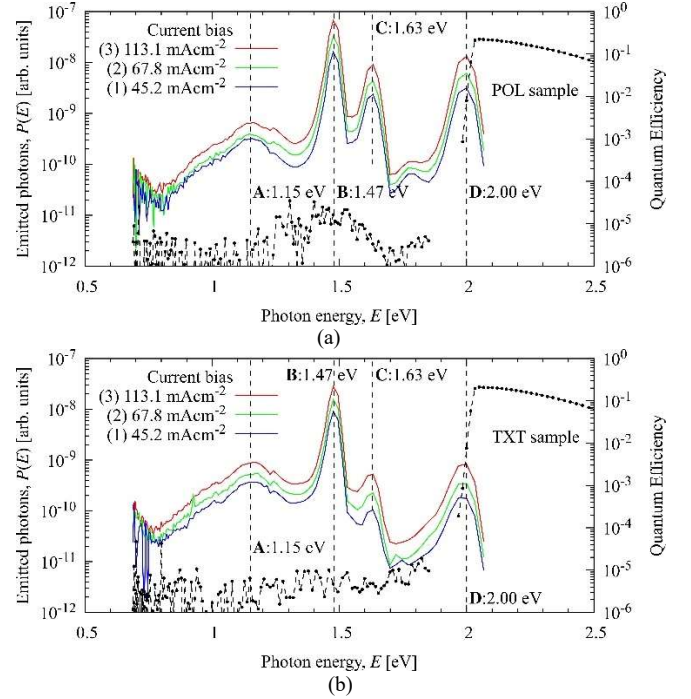

Fig. 5. Electroluminescent spectrum for the (a) POL sample and (b) TEX sample for different current bias conditions. Dots show the quantum efficiency for reference. The measurements are at 9K. The device junction area is 22.11 mm$^{-2}$.

ANNEX I: DETAILED SOLAR CELL STRUCTURE

The prototypes under study in this work consist of an AlGaAs p-i-n epitaxial solar-cell structure grown on a 360-μm-thick semi-insulated (SI)-GaAs wafer by molecular beam epitaxy (MBE) onto a 3" undoped (001) semi-insulated (SI)-GaAs wafers. Both the POL and TEX cells shown in this work belong to the same epitaxy, from which nine samples of each type were produced. Then, the samples were tested with probes and the best one of each group was chosen for mounting and wire-bonding. The corresponding results are the ones presented here. The rest were sacrificed so it must be taken into consideration that our results are based on a single sample of each group.

From bottom to top, the solar cells consist of the following layer structure (Fig. 6a):
- A highly n-type doped ($5 \times 10^{18} \text{cm}^{-3}$) 1,000 nm thick GaAs buffer layer, serving as well as n-type contact for sustaining lateral current towards the lateral contact.
- A 100 nm n + ($5 \times 10^{18} \text{cm}^{-3}$) back surface field (BSF) Al$_{0.41}$Ga$_{0.59}$As layer.
- A 200-nm-thick Si-doped Al$_{0.35}$Ga$_{0.65}$As (base) doped to $5 \times 10^{17} \text{cm}^{-3}$.
- One layer of InAs QDs embedded in an undoped layer of 600 nm-thick Al$_{0.35}$Ga$_{0.65}$As intrinsic layer. The QDs are self-assembled after the deposition of two monolayers of InAs. Slab doping with silicon is used to dope the dots with approximately one electron per dot. The QD areal density assumed is $3 \times 10^{10}$ cm$^{-2}$.

- An Al$_{0.35}$Ga$_{0.65}$As field damping layer (FDL) of 100-nm n-type ($3 \times 10^{16}$ cm$^{-3}$). The motivation of this layer is to drive the QDs as much as possible to a region where the electrostatic potential is flat [36].
- A 200 nm thick Al$_{0.35}$Ga$_{0.65}$As Be-doped emitter doped to $1 \times 10^{18}$ cm$^{-3}$.
- A 300 nm thick n$^{++}$ type ($1 \times 10^{19}$ cm$^{-3}$) Be doped GaAs contact layer.

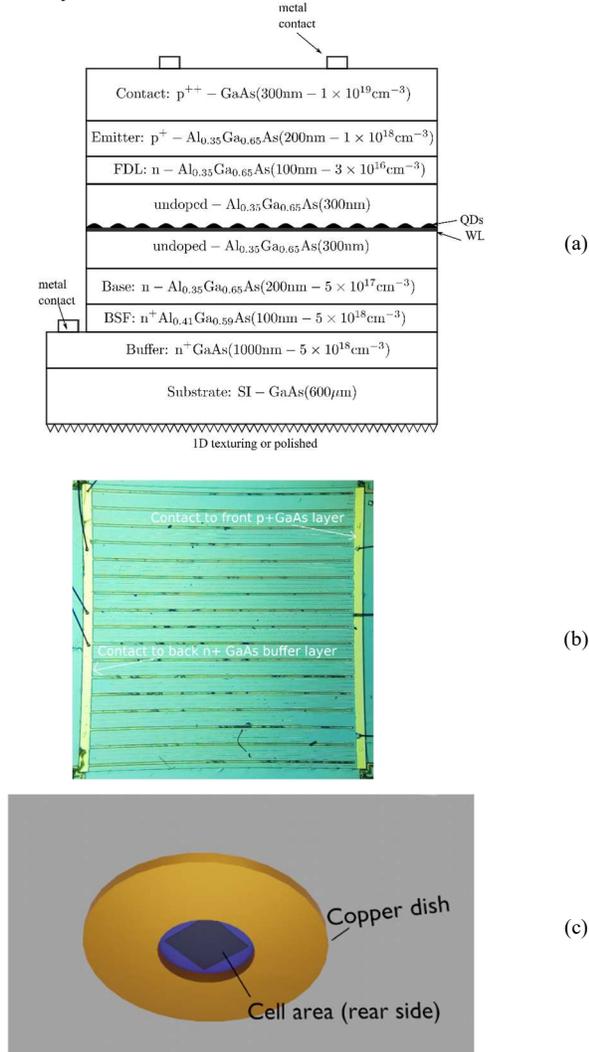

Fig. 6. (a) Detailed semiconductor layer structure of the solar cells used in this work; (b) optical photograph of the front size showing the grids that contacts the front and back of the cells. (Junction area, $A = 22.11$ mm²); (c) drawing of the assembly used to expose the rear of the cell to air.

Fig. 6b shows a picture of the front side of the cell. In this picture the interdigitated metal pattern that allows us to access both the $p$ emitter as to the $n^+$ buffer layer from the top can be appreciated. Fig. 6c shows schematically how the solar cells are assembled into a copper disc with a hole so that we can have air at the back of the solar cells. By having air we ensure that all the cells find the same medium (a medium with a refraction index of one) avoiding uncertainty in this respect from one cell to another.

ANNEX II: EXPERIMENTAL TECHNIQUES TO MEASURE THE OPTOELECTRONIC RESPONSE OF THE SOLAR CELLS

The photocurrent and quantum efficiency measurements were obtained using conventional lock-in techniques at 23 Hz chopping frequency and a Stanford Research low-noise transimpedance preamplifier. This preamplifier also served to bias the sample in short-circuit conditions during the measurements. The devices were illuminated with monochromatic light coming from a halogen lamp diffracted with a Newport ¼ monochromator for the case of visible and NIR ranges and a set of optical bandpass filters to remove secondaries. For the mid-infrared wavelengths, the halogen lamp is substituted by a SiC lamp. For the quantum efficiency measurements, Newport calibrated Si and Ge photodetectors were used. The absence of a calibrated detector for the mid-infrared range prevented us from transforming the photocurrent results in Fig. 4,a into quantum efficiency. Measurements at low temperature (9 K) were carried out inside a closed-cycle He cryostat.

The description of the electroluminescence measurement can be split into two parts: the excitation and the detection part. For excitation, a steady-state current was injected at a density from 45 to 113 mA/cm$^{-2}$ using a Keithley current source. For detection, luminescence from the sample was collected by a CaF$_2$ cell and directed via a set of mirrors into a 1/8 monochromator after it has passed through a mechanical chopper. A calibrated photodetector collects then the monochromatic light at the exit of the monochromator. The current response from the photodetector is amplified with the low-noise transimpedance preamplifier and measured using conventional lock-in techniques at a 23 Hz chopping frequency.

ANNEX III: THEORETICAL SIMULATION BASED ON RAY-TRACING

Ray-tracing, using the geometry sketched in Fig. 7,a, has been used to model the equivalent optical path of below bandgap energy photons, those which are weakly absorbed in the structure. The figure depicts a 300-µm GaAs ($n$=3.27) layer with a semi-circular one-dimensional pattern on the rear with the dimensions of the experimental pattern we have manufactured (Fig. 2,d). Our approach is approximated, since, for example we do not take into account the dependence of the refraction index with the photon wavelength nor diffraction effects. However, given the fact that the shortest relevant geometrical dimension is in the range of 4.5 µm, the ray-tracing approximation should be valid for photon wavelengths of $\lambda_0/n \lesssim 4.5$ µm, being $\lambda_0$ the photon wavelength outside the media, or photon energies of $\epsilon \gtrsim 0.08$ eV which indeed is the photon range of our experiments.

The structure is assumed to be surrounded by air ($n$=1). The simulation is carried out by launching rays within a cone with an aperture of 8º degrees on the front surface of the cell and, then, following the multiple reflections the ray suffers in its path as it encounters interfaces with a different refraction index between the media it separates. Fig. 7,b shows the results of the simulation in terms of the number *N*, which represents the number of times the ray crosses the plane containing the quantum dots. For example, we find that 25 % of the photons only cross this plane once (*N*=1). Given this distribution, we calculate that, on average, *N*=4.5. The same calculation of *N* is made for a flat structure (with equal properties and characteristics), resulting in *N*=1.32. This yields a relative increase in the mean number of ray-light passes of 3.4 inside the textured structure with respect to the polished one. This result is suitable to very low absorptance structures where the absorption is proportional to the number of light-ray passes. The 3.4 X relative increase is in agreement with our experimental results.

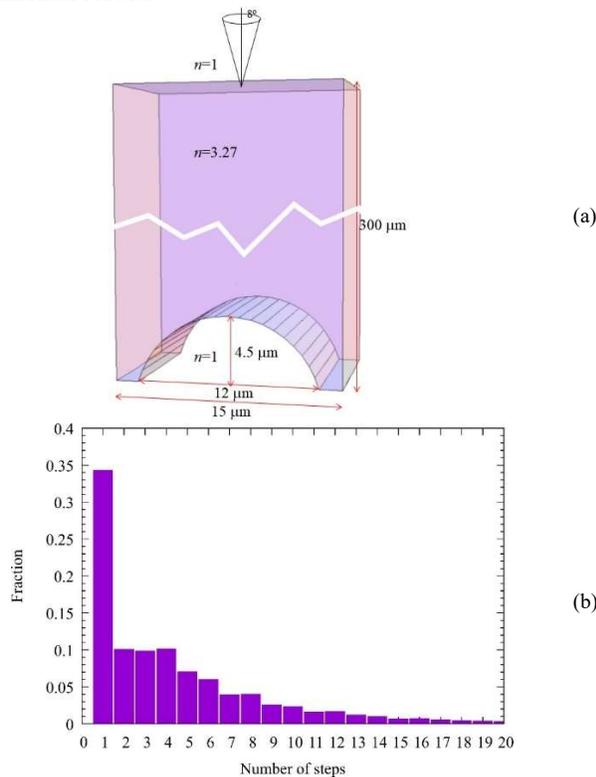

Fig. 7. (a) Unit cell structure used for ray-trace simulations. The refraction index, n, assumed in each of the regions is also indicated. (b) Distribution of the number of rebounds inside the structure.